\documentclass[pra,preprint,amssymb,showpacs]{revtex4}
\usepackage{dcolumn}
\usepackage{bm}
\usepackage{graphicx}
\def\be{\begin{equation}}
\def\ee{\end{equation}}
\def\bea{\begin{eqnarray}}
\def\eea{\end{eqnarray}}
\def\Eq#1{Eq.~\ref{#1}}
\def\Eqs#1#2{Eqs.~\ref{#1} and \ref{#2}}
\def\Ref#1{Ref.~\cite{#1}}
\def\Refs#1{Refs.~\cite{#1}}
\def\ie{{\it ie}}
\def\eg{{\it eg}}
\def\om{\omega}
\def\CI{{\cal I}}

\def\Tri{\hbox{Tr}^{(i)}}

\def\CO{{\cal O}}
\def\CP{{\cal P}}
\def\GG3{G^{(3)}}
\def\G4{G^{(4)}}
\def\C4{C^{(4)}}
\def\P4{P^{(4)}}

\def\ket#1{| #1 \rangle}
\def\bra#1{\langle #1 |}

\def\brakett#1#2#3#4{\langle #1,#2|#3,#4\rangle}
\begin{document}
\title{Entanglement Patterns in Mutually Unbiased Basis Sets
for N Prime-state Particles}
\author{Jay Lawrence}
\affiliation{Department of Physics and Astronomy, Dartmouth
          College, Hanover, NH 03755, USA}
\affiliation{The James Franck Institute, University of Chicago, 
          Chicago, IL 60637}
\date{revised \today}
\bigskip
\begin{abstract}
A few simply-stated rules govern the entanglement patterns that can occur in 
mutually unbiased basis sets (MUBs), and constrain the combinations of such 
patterns that can coexist (\ie, the stoichiometry) in full complements of ($p^N+1$) 
MUBs.  We consider Hilbert spaces of prime power dimension (as realized by  
systems of $N$ prime-state particles, or {\it qupits}), where full complements are 
known to exist, and we assume only that MUBs are eigenbases of generalized 
Pauli operators, without using a particular construction.   The general rules 
include the following:  1) In any MUB, a particular qupit appears either in a pure 
state, or totally entangled, and 2) in any full MUB complement, each qupit is pure 
in ($p+1$) bases (not necessarily the same ones), and totally entangled in the 
remaining ($p^N-p$).   It follows that the maximum number of product bases is 
$p+1$, and when this number is realized, all remaining ($p^N-p$) bases in the 
complement are characterized by the total entanglement of every qupit.   This
``standard distribution''  is inescapable for two qupits (of any $p$), where only 
product and generalized Bell bases are admissible MUB types.  This and the
following results generalize previous results for qubits  \cite{LBZ,Romero} 
and qutrits \cite{JL}, drawing particularly upon \Ref{Romero}.   With three qupits 
there are three MUB types, and a number of combinations ($p+2$) are possible 
in full complements.   With $N=4$, there are 6 MUB types for $p=2$, but new 
MUB types become possible with larger $p$, and these are essential to realizing 
full complements.  With this example, we argue that new MUB types, showing 
new entanglement characteristics, should enter with every step in $N$, and 
when $N$ is a prime plus 1, also at critical $p$ values, $p=N-1$.  Such MUBs 
should play critical roles in filling complements.
 \end{abstract}
\medskip
\pacs{03.65.-w, 03.65.Aa, 03.65.Ta, 03.65.Ud. }
\maketitle
\section{Introduction}

Mutually unbiased basis sets are known to provide an optimal basis
for quantum tomography \cite{Ivanovic,WWF}, to play key roles in 
quantum cryptography \cite{BB84,Ekert,BrussMac,Grob}, and to be 
instrumental in solving the mean king problem in prime power 
dimensions \cite{Aravind}.  The generalized Pauli operators associated 
with MUB's include the stabilizers of quantum error correcting codes 
\cite{Shor, Gottesman1, Calderbank}, and serve as entanglement 
witnesses \cite{Hyllus} for the MUB 
states.  Of interest for the foundations of quantum physics, the MUB 
concept sharpens the concept of complementarity \cite{BZ,LBZ}, and 
raises the question of existence in composite dimensions.  An excellent 
comprehensive review of MUBs has recently  appeared \cite{Durt}.

We deal here with Hilbert spaces of prime power dimensions ($d=p^N$), 
where $d+1$ MUBs are known to exist \cite{WWF}.  This is both the 
largest possible number, and also the number required for a complete 
operator basis (in representing the density matrix, for example).   So, while
each MUB is a complete orthonormal basis in the Hilbert space, the set of
$d+1$ MUBs is a complete (nonorthogonal) basis in the space of all
operators, which has dimension $d^2 = p^{2N}$.   Regarding terminology, 
to avoid reference to a ``complete set of complete sets,'' and prompted by
the fact that different MUBs (or the observable sets associated with them)
are maximally complementary \cite{BZ}, I will use the term ``full complement,'' 
or sometimes just ``complement,'' to denote the set of all $d+1$ MUBs.  
Partial MUB sets have been discussed in connection with composite 
dimensions and referred to as ``constellations'' \cite{Weigert}.

The natural systems to which MUBs apply consist of $N$ $p$-state objects 
({\it qupits}).   In such systems, while MUB complements exhibit only a single 
entanglement type for $N=2$ (and all $p$), the number of distinct types 
proliferates with increasing $N$.  The variety is illustrated in a number of 
recent discussions, mostly on multiple qubit systems but also multiple qutrit 
systems \cite{LBZ,JL,Romero,BR1,Grob,Wiesniak,Klimov}.  In particular, 
a systematic study by Romero and collaborators \cite{Romero} illustrates a 
broad range of entanglement patterns that occur naturally in a construction 
scheme for full MUB complements.   Such complements are catalogued for 
up to 4 qubits.  Wie\'{s}niak and collaborators \cite{Wiesniak} have developed 
a construction scheme aimed at experimental implementation and discussed
the total entanglement content of full MUB complements of bipartite systems. 

With the general MUB problem in mind, our purpose here is to develop a 
general framework, independent of construction schemes, for exploring MUB 
entanglement patterns for all $p$ and $N$.  The project begins by proving 
three general theorems (the ``rules'') that underlie and lead quickly to an 
array of more specific results.
Many of the latter apply to all $p$, but are $N$-specific, as each step in 
$N$ introduces further richness.  All results refer to one of two levels - that 
of individual MUBs and that of full complements.    At the individual level, 
MUB types are characterized by first specifying the {\it separation pattern} -  
How many, and how big, are the irreducible subsets of qupits defined by the 
factorization of the wavefunction? - and next, by describing the 
{\it entanglement pattern} - What is the nature of the entanglement within each 
irreducible subset?   At the level of the full complement, we ask about the
possible {\it MUB distributions} - What combinations of MUB types can coexist 
within full complements.   At the first level, we will show that all conceivable 
separation patterns are possible, and we will show with examples how to 
describe the entanglement within the nonseparable factors.   At the level of the 
full complement, we will show how to deduce constraints on the possible MUB 
distributions.    For $N=2$ and 3, surprisingly, the general global constraints 
mentioned in the abstract suffice to determine all MUB distributions for all $p$.
The $N=4$ case is considerably more complex and requires the derivation of 
more detailed constraint equations.

Let us begin with a review of basic concepts and notation in Section II.  In 
Section III we prove the three general theorems.  These rules are applied 
in Section IV to obtain the entanglement patterns of individual MUBs, 
and to deduce constraints on their possible distributions within full 
complements, taking the $N=2$ - 4 cases in turn.  In Section V we 
summarize results and comment on unresolved questions.
\section{Background Concepts and Definitions}
In Hilbert spaces of dimension $d$, two orthonormal bases ($K$ and $Q$) 
are mutually unbiased if any state $\ket{K,k}$ in basis $K$ has uniform
probably of being found in any state $\ket{Q,q}$ in basis $Q$; that is, if
\be
  |\brakett{K}{k}{Q}{q}|^2 = 1/d.
\label{MUB}
\ee
Thus, measurements in the two bases provide no redundant information.  
Since measurements in any basis provide $d-1$ independent
probabilities, and since $d^2-1$ real parameters are needed to determine an 
unknown quantum state (its density matrix $\rho$),  it follows that $d+1$ MUB's 
are required.  In this way the MUB projectors form a complete nonorthogonal 
basis in operator space.  This required number of MUBs is (only) known to 
exist in power-of-prime dimensions.  
 
There is an intimate connection between MUBs and generalized Pauli operators 
(hereafter called simply ``Pauli operators'') which underlies several construction 
schemes (\Ref{Durt} provides a comprehensive listing \cite{Durt2}).   These 
operators are conventionally written in the form of a tensor product,
\be
    \CO_{n,m} = X^nZ^m \equiv 
    X_1^{n_1}X_2^{n_2}...X_N^{n_N}Z_1^{m_1}...Z_N^{m_N},
\label{Pauli1}
\ee
whose factors, acting on individual qupits, are powers of the generalized 
($p \times p$) Pauli matrices,
\be
  Z = \sum_{k=0}^{p-1} \ket{k} \om^k \bra{k} \hskip1truecm
  \hbox{and} \hskip1truecm X = \sum_{k=0}^{p-1} \ket{k+1} \bra{k},
\label{Pauli2}
\ee
where $\om = e^{2\pi i/p}$, and $X$ is the raising operator of $Z$.   The powers $n$
and $m$ are $p$-nary numbers, {\it eg}, $n=(n_1,...n_N)$, whose digits take the 
values 0,1,...,$p-1$.  Thus, there are $p^{2N}$ operators $\CO_{n,m}$  (including 
the identity $\CI = \CO_{o,o}$), which make up a complete and orthonormal basis
in operator space (with the trace operation as inner product).   The desired connection 
with MUB's is described in \Ref{Bandyo}:  The $\CO_{n,m}$ partition into $p+1$ 
internally-commuting subsets, each consisting of $p-1$ traceless operators (excluding 
$\CI$).  The corresponding eigenbases then form a complete complement of MUB's. 

The above are standard definitions and conventions.  It will be useful to adopt 
a couple of more special conventions for use throughout this paper.  First, the 
operator set $\CO_{n,m}$ does not form a group, because multiplication 
generates irreducible phase factors.   However, for odd $p$ the set 
$\CO_{n,m} \otimes (1,\om,...,\om^{p-1})$ does form a group, of order $p^{2N+1}$, 
and for $p=2$ the analogous set $\CO_{n,m} \otimes (\pm 1, \pm i)$ forms a group 
of order $2^{2N+2}$.  These are called discrete Heisenberg-Weyl, or generalized 
Pauli groups \cite{Durt,NC}.  We shall not make direct use of them, but we shall take 
advantage of the freedom to redefine the phases of the $\CO_{n,m}$ in the original 
set:  We choose phases so that the compatible subsets form groups, and we call 
these {\it compatibility groups}.  They are all isomorphic to those consisting of 
$X_{}^n$ and $Z_{}^m$, each of which is generated by the $N$ independent 
elements, $X = (X_1,...,X_N)$ and $Z = (Z_1,...,Z_N)$, respectively.  Thus, to 
construct another compatibility group, we may choose a generator set 
$G = (G_1,...,G_N)$ that consists of any $N$ elements in the original compatible 
subset that do not form a subgroup, and write the resulting group elements as 
\be
  G^n \equiv G_1^{n_1}G_2^{n_2}...G_N^{n_N}.
\label{GGroup}
\ee 
Thus, all of the compatibility groups are representations of the same
group - the abelian group of order $p^N$ generated by $N$ elements.
A simple example of a compatibility group so generated is 
\be
   Y^n = (X_1Z_1)^{n_1}(X_2Z_2)^{n_2}...(X_NZ_N)^{n_N}.
\label{YGroup} 
\ee
Note that phase factors are introduced with respect to the original Pauli 
operators because, \eg, $(X_iZ_j)^2 = \om^{\delta_{ij}}  X_i^2Z_j^2$.

The generator set $G$, by itself, completely determines the states of the 
basis ($G$) in the Hilbert space, through the eigenvalue equations 
$G_i \ket{G,k} = \om^{k_i} \ket{G,k}$, where $k = (k_1,k_2,...,k_N)$ 
is a $p$-nary representation of the state index $k$.  The eigenvalues 
of a general group element are then given by
\be
   G^n \ket{G,k} = \om^{n \cdot k} \ket{G,k},
\label{eigenvalues}
\ee
where $n \cdot k = n_1k_1 + n_2k_2 + ... + n_Nk_N$, and the spectral 
representation of $G^n$ is therefore just the Fourier transform \cite{transform}
\be
   G^n = \sum_k \ket{G,k} \om^{n \cdot k} \bra{G,k} \equiv 
   \sum_k \om^{n \cdot k} \CP(G,k),
\label{specrep}
\ee
where $\CP(G,k)$ is the projector onto state $k$ in basis $G$.  This MUB 
projector is then given by the inverse transform,
\be
   \CP(G,k) = p^{-N} \sum_n \om^{-n \cdot k} G^n.
\label{inverse} 
\ee
The existence of these simple transform relationships between every 
compatibility group and its corresponding MUB projector set is a
consequence of defining the former to be a group.  The only remaining
arbitrary phases are those of the generators.

\section{General Results on Entanglement}
In this section we establish the general rules that will form the basis for 
the rest of the work.   For ease of reference and completeness I will state and 
prove these results as three separate numbered theorems.   For transparency, 
here, in plain English, is what they will say about MUB states:  (I)   A given qupit 
is perfectly pure or totally entangled,  (II)  The distribution of one-qupit operator 
factors in the compatibility group correlates with this purity,..., and (III)  In any full 
MUB complement, every qupit appears pure $p+1$ times, and totally entangled 
$p^N-p$ times.

These theorems and the results that follow from them rely on the assumption 
that MUB states are eigenstates of Pauli operators.  While this is restrictive for 
individual MUB pairs, it is not restrictive for known MUB complements or known 
construction schemes \cite{Durt2, Boykin}, allowing for unitary equivalence.   
An example may help to illustrate.  
Consider the standard basis in 4D, and another related to it by the unitary 
transformation $U_{n,m} =  (i)^{nm}/2$ (where $n,m = $ 0,1,2,3), which is not an 
eigenbasis of the Pauli operators of \Eq{Pauli1}.   The two bases are MU, but a full 
complement cannot be completed containing both of them.   However,  full 
complements can be found containing either basis without the other:  Starting with 
the well-known full complement containing the standard basis,  one could apply 
$U_{n,m}$ to each of its bases to obtain another full complement.   The 
latter are not eigenbases of the original Pauli operators, but clearly they {\it are} 
eigenbases of transformed Pauli operators, which may be thought of as 
corresponding to redefined parts (and redefined quantization axes).  The results 
of this paper then apply with reference to these redefined parts.   Regarding the 
existence of a MUB complement outside of this equivalence  - I believe that this
question also remains unresolved \cite{Boykin}.   We will return to these points in 
the conclusions.

As a brief preliminary, one-qupit states within the $N$-qupit system are defined 
by the reduced density matrices,
\be
    \rho_i = \Tri \rho,
\label{rhoi}
\ee
where Tr$^{(i)}$ denotes the partial trace over states of all but the $i$-th qupit.   
Perfect purity means that $\rho_i = \rho_i^2$ is a projector, while total impurity 
means that $\rho_i = \CI/p$.   One can define the purity of the state $\rho_i$ as
\be
  P_i = (pTr \rho_i^2 - 1)/(p-1), 
\label{purity}
\ee
which takes its extremal values, 1 and 0, in the respective cases. 

\noindent {\bf Theorem I:} If the system is in a pure eigenstate of 
Pauli operators (a generator set $G$), then any individual qupit 
must exist in a state of either perfect purity, or total impurity, the 
same for all eigenstates of $G$.

\noindent {\bf Proof:} The generators produce a compatibility group, 
and the $N$-qupit density matrix representing a pure eigenstate, 
$\rho = \CP(G,k)$, may be expanded as in \Eq{inverse}.  Considering 
now the Pauli matrix factors that act 
on just the $i$th qupit, the generator set $G$ must fall into one of two 
categories:  Only one Pauli matrix, say $Z_i$ (and possibly powers of it), 
appears in the generator set, or more than one appear (including, say,
$X_i$ and $Y_i$), that are not powers of one another.  Consider 
the latter case, which is simpler:  Let $G_1$ and $G_2$ be generators 
that contain the factors $X_i$ and $Y_i$.   No operator of the form 
$U_i \CI$ (where $U_i$ is any one-qupit Pauli matrix) commutes with both 
$G_1$ and $G_2$, and all such operators are thereby excluded from the 
compatibility group.  As a result, the only operator with a 
nonvanishing partial trace Tr$^{(i)}$ is the global identity $\CI$.  
Since $\CI$ enters the summation (\ref{inverse}) with the coefficient 
$p^{-N}$, and Tr$^{(i)}$ produces a factor of $p^{(N-1)}$, the reduced 
density matrix for the $i$th qupit is 
\be
  \rho_i = p^{-1}~\CI_i,
\label{T1}
\ee
indicating that the $i$th qupit is totally impure.

Now turn to the other case:  If only $Z_i$ (and possibly powers) appear 
in the generator set, then only $Z_i$ and its powers can appear in the 
compatibility group (again refering only to those factors that act on the 
$i$th qupit.  Since the ``one-body'' operators $Z_i^{n_i} \CI$ commute 
with all of these, they must belong to the compatibility group.  These 
one-body operators are the only ones that survive the partial trace.  
Since each of them enters the summation (\Eq{inverse}) with 
coefficient $p^{-N}\om^{-n_ik_i}$, and since $Tr^{(i)}$ produces 
a factor of $p^{(N-1)}$ in each term, we find in this case that 
\be
  \rho_i = p^{-1} \sum_n \omega^{- n_ik_i} Z_i^{n_i} = 
  \ket{Z_i,k_i} \bra{Z_i,k_i}.
\label{T2}
\ee
This shows that $\rho_i$ is a projector onto the eigenstate of 
$Z_i$ whose eigenvalue is $\omega^{k_i}$, that is,
\be
   \rho_i^2 = \rho_i  \hskip1.2truecm \hbox{and}
    \hskip1.2truecm    Z_i \rho_i = \omega^{k_i} \rho_i.
\label{T3}
\ee
This proof is independent of the choice of the eigenstate 
$k = (k_1...k_N$) in the basis $G$, and so clearly the $i$th qubit 
is perfectly pure for all eigenstates in this basis .  

Here is a related more detailed theorem on the distribution of 
one-qupit matrices associated with a single qupit.

\noindent {\bf Theorem II:} In any compatibility group of $N$-qupit 
Pauli operators, the distribution of one-qupit factors acting on 
the $i$th qupit must be one of two types:  (i) Only a single Pauli 
matrix and its powers occur, and each power occurs an equal 
number ($p^{N-1}$) of times, or (ii) every Pauli matrix occurs, 
and each occurs an equal number ($p^{N-2}$) of times.

\noindent {\bf Proof:} Consider any set $G$ of $N$ generators of 
the compatibility group.  This set must be one of the two types
considered in the foregoing proof:  Suppose first that only one 
Pauli matrix (say $Z_i$), and possibly powers of $Z_i$ appear.  
Let $G_1$ be a generator containing $Z_i$ as a factor, and let $G_2$, 
$G_3$,..., $G_N$ be the rest.  $G_1$ by itself generates a cyclic 
subgroup containing all powers of $Z_i$.  Then, $G_1$ and $G_2$ 
by themselves generate a subgroup of order $p^2$ in which, by 
virtue of the rearrangement theorem, every power of $Z_i$ appears 
$p$ times (no matter which power of $Z_i$ is present in $G_2$).  
One may repeat this argument, multiplying the order of the subgroup 
by $p$ at each stage, until the full compatibility group is generated, 
with each power of $Z_i$ being produced $p^{N-1}$ times.

In the other case, let $G_1$ and $G_2$ be generators containing
the $X_i$ and $Y_i$ factors, respectively.  These two generators,
by themselves, generate a subgroup of order $p^2$ in which every
Pauli matrix factor $U_i$ appears once and only once.  (To see
this, note that $X_i$ and $Y_i$, by themselves, generate the 
one-qupit Pauli group \cite{Pauligroup}, but since $G_1$ and 
$G_2$ commute, the multiplicity of phase factors is absent.)  Now, 
by including a third generator, $G_3$, one generates a subgroup 
of order $p^3$ in which, by the rearrangement theorem, each 
Pauli matrix factor appears $p$ times. Repeating the process 
through $G_N$, one generates the full compatibility group with 
each Pauli matrix factor appearing $p^{N-2}$ times.

The second result is particularly striking in light of the fact
that the nature of the entanglement of the $i$th qupit may vary
widely, in the sense that its entanglement may be shared with any 
number of other qupits in the system.  Nevertheless, only two kinds
of Pauli matrix distributions, with the correponding purities, are
possible. 

We use both of the foregoing theorems to deduce the total
entanglement content - as measured by the one-qupit purities  - of a 
full complement of MUB's. This total content is constrained by the 
requirement that the two types of one-qupit Pauli matrix distributions 
be consisent with the set of all Pauli operators, which must appear 
in the full complement. 

\noindent {\bf Theorem III:} Within any full complement of $p^N+1$
MUB's, every qupit is perfectly pure in $p+1$ basis sets, and
totally entangled in the remaining $p^N-p$.

\noindent {\bf Proof:} Consider the $i$th qupit.  Recall that the 
total number of Pauli operators (excluding $\CI$) is $p^{2N}-1$,
and that these exactly accommodate the $p^N+1$ compatibility 
groups containing $p^N-1$ traceless operators each. Each Pauli 
matrix factor $U_i$ appears in $p^{2N-2}$ Pauli operators, except 
for $\CI_i$ which appears in $p^{2N-2}-1$ because we are not 
counting $\CI$ in the individual groups. This number must equal 
the sum of $\CI_i$ factors appearing in all of the compatibility 
groups.  According to the previous theorem, there are $p^{N-1}-1$ 
such factors in compatibility groups in which the $i$th qupit is 
pure, and $p^{N-2}-1$ such factors in all other compatibility
groups.  If $\nu_S^{}$ is the number of 
compatibility groups (or basis sets) in which it is pure, then, in
order to account for all $\CI_i$ factors, we must have
\be
  p^{2N-2} -1 = \nu_S^{} (p^{N-1} -1) 
  + (p^N+1-\nu_S^{})(p^{N-2} -1).
\label{accounting}
\ee
Solving this equation, we find the number of basis sets in
which the $i$th qupit is pure,
\be
  \nu_S^{} = p+1,
\label{nusubS}
\ee
and consequently, the number of basis sets in which it is 
totally entangled,
\be
  \nu_E^{} = p^N - p.
\label{nusubE}
\ee

The following corollary arises when all qupits take their pure states 
simultaneously:
\noindent {\bf Corollary}:  The maximum number of product MUBs is $p+1$, and 
in any MUB complement where this number is realized, all of the remaining 
MUBs ($p^N-p$) must be totally entangled (in the sense that every qupit is 
totally entangled)  \cite{Wiesniak2}.  This is the standard distribution.  

Note that the probability of finding the $i$th qupit pure in a MUB state picked
at random from any full complement is equal to the averaged purity (\Eq{purity}),
\be
     \langle P_i \rangle_{comp}^{} = {\nu_S^{} \over \nu_S^{}+\nu_E^{}} = 
     {p+1 \over p^N+1},
\label{Piave}
\ee
which vanishes exponentially with $N$.   

\section{Entanglement Patterns and their Stoichiometries}
We discuss the $N=2 - 4$ cases in turn.  The first two are simpler, and we find
that Theorems I and III are sufficient to determine all possible MUB distributions,
although II provides useful insights.  With $N=4$, we require Theorem II in 
deriving more detailed constraints that apply to individual qupits.   

\smallskip
\centerline{{\bf bipartite systems}}

Clearly, if one qupit is pure, then so must be the other.  In light of Theorem I, 
then, both purities must be unity, or both zero.   Because these purities coincide,
the corollary of Theorem III applies:  There are $p+1$ product bases and $p^2-p$ 
totally entangled bases - the standard distribution is inevitable.  

We shall refer to all of the entangled bases as generalized Bell bases, because 
they share the common property that their compatibility groups consist 
solely of two-body operators, \ie, those containing no $I_k$ factors \cite{pure}.    
To see the consequences of this, write one of the two generators as $G_1 = UV$.  
The most general eigenstates of $G_1$ may then be written as  $p$-term 
expansions in the product basis of $IV$ and $UI$, 
\be
   \ket{\psi} = {1 \over \sqrt{p}} \sum_k C_k \ket{k}_u \ket{q-k}_v,
\label{Bform}
\ee
where the eigenvalues of $UV$ are $\om^q$ and the coefficients $C_k$ are 
determined by the other generator, call it $G_2 = ST$.   Commutativity demands 
that both $S \neq U$ and $T \neq V$,  so $G_2$ induces cyclic permutations (of 
order $p$) in the product states $\ket{k}_u \ket{q-k}_v$.   Therefore the $C_k$ 
are unimodular, and the $p$ eigenvalues ($\om^r$) of $G_2$ are nondegenerate, 
like those of $G_1$.   This confirms explicitly what we know from Theorem I - 
namely, that measurements of one-qupit properties (\eg, $IV$ or $UI$) must 
produce random distributions over all possible outcomes.  

The generalized Bell states defined above are contained within a broader class 
definitions given elsewhere \cite{Klimov2,Durt04}.   The more restrictive definition 
given here - defining classes of states by the Pauli operators of which they form 
eigenbases - applies nonetheless to all MUBs that are compatible with known full 
complements, and we shall employ such definitions throughout this work as we 
proceed to larger N.  

We note for future reference that the precise form of the product state expansion 
(\ref{Bform}) depends on the choice of basis.  A bad choice would require a 
$p^2$-term expansion, but even a good choice could look slightly different.  For 
example, if eigenstates of $UV^{-1}$ were expanded in the same product basis 
used in \Eq{Bform}, one would find sums of $\ket{k}_u \ket{q+k}_v$.  

As a final note on Bell states, our working definition may be given in words alone:  
A generalized Bell state is any totally entangled two-qupit eigenstate of Pauli 
operators  (since total entanglement requires that the two Pauli operators be 
of the form $UV$ and $ST$).

\smallskip
\centerline{{\bf tripartite systems}}

The standard MUB complement has $p+1$ product bases and $p^3-p$ 
totally entangled bases.   We shall refer to all of {\it these} totally entangled
bases as generalized GHZ, or $G$-bases, because they have common 
properties describable as follows:  

Let us first illustrate with a specific example that generalizes a standard 
choice of generators for qubits \cite{Mermin}, 
\be
   G \equiv (G_1, G_2, G_3) = (XXY, ~XYX, ~YXX),
\label{GHZ1}
\ee
to arbitrary $p$.   To identify an optimal product basis for an expansion, 
replace the latter two generators by 
$G_2' = G_2G_1^{-1}$ and $G_3' = G_3G_1^{-1}$.   
Recalling the usual definition $Y_i =X_i Z_i$ on the $i$th qupit (modulo 
possible phase factors), the result is
\be
   G' = (XXY,~IZZ^{-1},~ ZIZ^{-1}).
\label{GHZ2}
\ee
Clearly the most general joint eigenstates of $G_2'$ and $G_3'$ are  
$p$-term expansions in the standard basis,
\be
   \ket{\psi} =  {1 \over \sqrt{p}} \sum_k C_k \ket{k+q} \ket{k+r} \ket{k},
\label{GHZ3}
\ee
where $\om^r$ and $\om^q$ are the eigenvalues of $G_2'$ and $G_3'$, 
respectively, and the $C_k$ are determined by $G_1$.   The $C_k$ are 
again unimodular because $G_1$ generates a cyclic group of order $p$, 
of which the $p$ product states form a basis.   This again illustrates the 
randomness of one-qupit properties in totally entangled states.

To demonstrate the commonality of all totally entangled three-qupit bases,
we note that at least one generator must be a three-body operator  (having
no $I_k$ factors), which we write in complete generality as $G_1 = UVW$.   
Now, according to Theorem II, the inverse of each factor occurs $p$ times 
in the compatibility group, once with the inverse of $G_1$ itself, and $p-1$ 
times in other three-body operators in which it is the only inverse (footnote 
\cite{pure}).    Choosing two from the latter category, one containing $U^{-1}$ 
and the other containing $V^{-1}$, and multiplying $G_1$ by each in turn,
we obtain the generator set 
\be
   G = (UVW,~IBC,~AIC),
\label{GHZ4}
\ee
where compatibility requires that $C$ is common to $G_2$ and $G_3$
as indicated.   Clearly, $A$, $B$, and $C$ define the product basis for the
$p$-term expansions,
\be
 \ket{\psi} =  {1 \over \sqrt{p}} \sum_k C_k \ket{q-k}_a \ket{r-k}_b \ket{k}_c,
\label{GHZ5}
\ee
and each of $A$,  $B$, and $C$ must differ from corresponding factors
that appear in three-body operators of the compatibility group.  In other 
words, every three-body operator in the compatibility group induces cyclic 
permutations of the states composing the product basis.   The similarity
of generator sets shows that all totally entangled three-qupit bases have
$p$-term expansions in some special product basis, and that all of their
compatibility groups (of the same $N$ and $p$) have the same numbers 
of three-body and two-body operators.

Again, a purely verbal definition is possible:   A generalized GHZ state is any 
totally entangled 3-qupit eigenstate of Pauli operators.   A general statement
for $N \geq 4$ is possible but less categorical. 

The new aspect of MUBs that enters with $N=3$ is the appearance of a
third (nonstandard) MUB type, and with it,  the possibility of composing a full
complement with varying combinations.   The third type is biseparable, and
thereby nonsymmetric with respect to qupits - one qupit separates, leaving 
the other two in a Bell state.   We shall refer to these as ``separable-Bell'' 
bases, with the shorthand notation $SB$ (or $S_i B$ if we wish to identify 
the pure qupit).  Such MUB bases are known for $p=2$ and 3 (\Refs{LBZ,JL}),
and to describe them for arbitrary $p$,  we consider a generator set
\be
   S_1B = (IIA,~UVI,~STI), 
\label{SB1}
\ee
where $UV$ and $ST$ are commuting two-body operators acting on qupits 
2 and 3.   The $p^3$ joint eigenstates of this set may be written as
\be
   \ket{S_1B:k,q,p} = \ket{A_1:k}\ket{B_{2,3}:q,p},
 \label{SB2}
 \ee
 which describes qupit 1 in the $k$th eigenstate of $A$, and qupits 2 and 3 in 
 the Bell state denoted by the eigenvalues $q$ and $p$ of $UV$ and $ST$, 
 respectively.   Similarly, the compatibility group of $S_1B$ is a tensor product 
 of that associated with qupit 1 ($I_1,A_1,...,A_1^{p-1}$) and that of the Bell 
 basis of qupits 2 and 3.    The tensor product is a common characteristic of all 
 separable MUBs, and the eigenstates of a particular MUB all have the same 
 character - the separation pattern involves the same entangled subsets of 
 qupits, and the nature of the entanglement within each subset is the same.
 
 The three MUB types discussed above, including the three variations of the 
 SB bases, exhaust all of the possibilities for three qupits.

The remaining question now is, what combinations the three types of bases 
may appear in the full complement?  One can answer this question simply by 
conserving the number of pure qupits while conserving the number of basis 
sets.   We then find that we can remove a single product basis ($\Pi$) while 
adding three $SB$ and removing two $G$ bases:
\be
   \Pi + 2G \rightleftharpoons 3SB
\label{stoich1}
\ee
Table I shows the possibilities for three particles with any $p$.  The cases of
$p = 2$ and 3 dramatize the role of totally entangled states with increasing 
dimension of the Hilbert space.  In fact, case (a), dimension $d=8$, is the 
only multiparticle MUB dimension in which a complement can be found with 
no totally entangled bases.    And more typically, a majority of MUBs are totally 
entangled:  In case (b) at least 4/7 of all bases are $G$ bases, and even for 
{\it two} qutrits, 6 of the 10 bases are Bell bases.     For $N=3$ and general $p$, 
the minimum number of $G$ bases is given by $N_{min}(G) = p^3 - 3p - 2$, 
an ever-increasing fraction of the total number of bases as $p$ increases.
\begin{table}
$$
\vbox{
  \halign{
   \hfil # \hfil \quad & \hfil # \hfil \quad & \hfil # 
   \hfil \quad & \hfil # \hfil \quad & \hfil #
   \hfil \quad & \hfil # \hfil                            \cr
\noalign{\hrule}
\noalign{\smallskip}
     & \multispan{3 (a) three qubits}    \cr
\noalign{\smallskip}
\noalign{\hrule}
\noalign{\smallskip}
    $\Pi$ & 3 & 2 & 1 & 0 &         \cr
    $SB$ & 0 & 3 & 6 & 9 &          \cr
    $G$ & 6 & 4 & 2 & 0 &       \cr
\noalign{\medskip}
\noalign{\hrule}
}}  
\hskip1.0truecm
\bigskip
\vbox{
  \halign{
   \hfil # \hfil \quad & \hfil # \hfil \quad & \hfil # 
   \hfil \quad & \hfil # \hfil \quad & \hfil #
   \hfil \quad & \hfil # \hfil                            \cr
\noalign{\hrule}
\noalign{\smallskip}
     & \multispan{3 (b) three qutrits}    \cr
\noalign{\smallskip}
\noalign{\hrule}
\noalign{\smallskip}
    $\Pi$ & 4 & 3 & 2 & 1 & 0         \cr
    $SB$ & 0 & 3 & 6 & 9 & 12         \cr
    $G$ & 24 & 22 & 20 & 18 & 16      \cr
\noalign{\medskip}
\noalign{\hrule}
}}  
$$                            
\end{table}
\begin{table}
$$
\vbox{
  \halign{
   \hfil # \hfil \quad & \hfil # \hfil \quad & \hfil # 
   \hfil \quad & \hfil # \hfil \quad & \hfil # \hfil      \cr
\noalign{\hrule}
\noalign{\smallskip}
     & \multispan{1 (c) three qupits}    \cr
\noalign{\smallskip}
\noalign{\hrule}
\noalign{\smallskip}
    $\Pi$ & $p+1$ & $p$ & ... &  0         \cr
    $SB$ & 0 & 3 & ... & $3(p+1)$        \cr
    $G$ & $p^3-p$ & $p^3-p-2$ & ... & $p^3-3p-2$      \cr
\noalign{\smallskip}
\noalign{\hrule}
\noalign{\smallskip}
\noalign{\hrule}
}}  
$$                            
\caption{Numbers of product, separable-Bell, and GHZ bases coexisting
for three particles.}
\end{table}

It is noteworthy that $SB$ bases can be introduced only in steps of three, 
reflecting the condition that the three variations $S_iB$ must balance in the 
full complement, since the other MUB types are symmetric with respect to
permutations of qupits.  This condition follows from the conservation of pure 
states for each qupit separately.

\centerline{{\bf quadrapartite systems}}

The $N=4$ case is more complex in a number of respects.   Most importantly, new 
MUB types enter with increasing $p$.    But even with $p=2$, the number of distinct 
MUB types exceeds the number of separation patterns.   Figure 1 shows the five 
separation patterns that characterize all $p$, and lists seven MUB types, six of 
which account for all $p=2$ options, and a seventh which represents, but is not
exhaustive for $p \ge 3$.   Let us first discuss the MUB types for general $p$, and 
later specialize to particular cases for constraints and stoichiometries.
\begin{figure*}[ht]
\centerline{
\includegraphics[width=6cm,angle=90,clip]{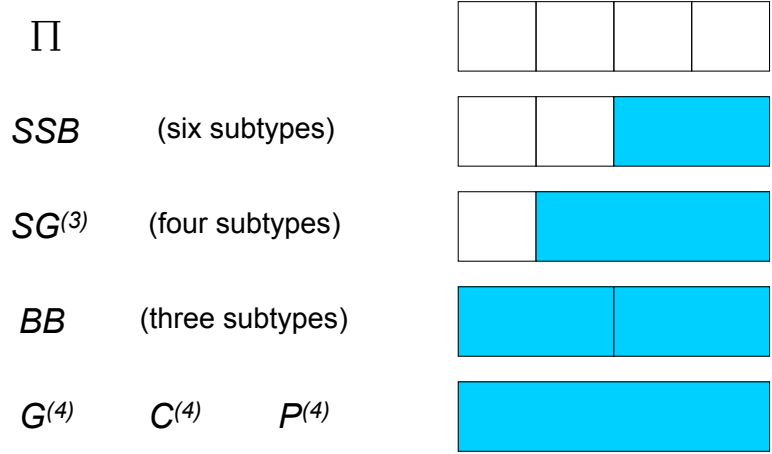}}
\caption{\label{fig1} Seven MUB types listed with 5 separation patterns for 4 qupits.} 
\end{figure*}

The separable MUBs' compatibility groups are tensor products of those of their
constituent MUBs, and their generator sets may be constructed accordingly.
With $SG^{(3)}$, a single generator is associated with the separating particle 
(for example $IIIA$), while three generators (each of the form $UVWI$ or their 
alternatives) are associated with the three particles forming $G^{(3)}$ states.  
There are four variations on this pattern corresponding to the choices of the
separating particle.   In the $BB$ case, one could pick two generators of the form 
$IIUV$, and two of the form $STII$.   There are three variations on this separation
pattern, corresponding on the three ways of picking the two entangled pairs, as
compared with six variations on the $SSB$ pattern from the six ways of picking
a single entangled pair. 

Let us discuss the nonseparable bases in somewhat more detail beginning with  
four-particle GHZ bases ($G^{(4)}$).  These are straightforward generalizations 
of the three-particle bases $G^{(3)}$, and a standard generator set \cite{Mermin}
consists of the four operators
\be
  G^{(4)} =  (XXXY,~ XXYX,~XYXX,~YXXX). 
\label{genG}
\ee
From an alternative generator set, ($XXXY,~IIZZ^{-1},~IZIZ^{-1},~ZIIZ^{-1}$), it is 
apparent that eigenstates may again be written as superpositions of $p$ product 
states in the standard basis.   A more general characterization of GHZ states
is provided in the Appendix.

Cluster bases ($C^{(4)}$) were introduced in connection with measurement-based,
one-way quantum computation  \cite{BR2}, and in fact both cluster and GHZ states 
are special cases of a broad class of $N$-qubit states, called graph states, which 
form  the basis of this \cite{BR3}.  \Ref{Klimov} has shown that graph states may be 
classified in terms of curves in phase space, which provides a further connection 
with the MUB problem.  Cluster bases are defined here, for all $p$, by generator 
sets of which a standard example, introduced for the qubit case \cite{BR2}, is
\be
  C^{(4)} = (XZXI,~ ZXIX, ~XIXZ, ~IXZX).
\label{genC}
\ee
Cluster states have stronger entanglement links between smaller groupings of 
particles, making their entanglement more robust against decoherence \cite{BR1} 
than GHZ entanglement, which is shared equally among all particles.  This is 
reflected in the fact that $\C4$ has only two 2-body operators in its compatibility
group, as compared with three in the $\G4$ case.  For this reason, its generator set 
can only be simplified to  ($XZXI,~ZIZ^{-1}I,~IZ^{-1}IZ,~IXZX$), and as a result, the 
eigenstate expansions can be reduced to no less than $p^2$ terms in the standard 
basis.   A general characterization of  $C^{(4)}$ accompanies that of $\G4$ in the 
Appendix, which then goes on to show that these, together with the four separable 
bases, exhaust all MUB possibilities for four qubits.   

As a final example, I have found that a new type of basis, one that has no counterpart 
for qubits, is necessary for the existence of full MUB complements when $p>2$, for 
reasons that will become apparent.    A generator set  giving rise to such a basis is
\be
    P^{(4)} = (ZXYW,~XZWY,~WYXZ,~YWZX),
\label{genP}
\ee
where standard definitions $Y=XZ$ and $W=XZ^2$ are followed.   The essential 
point is that the generators are tensor products of four noncommuting one-body 
matrices, which rules out qubits, but makes possible the elimination of 2-body 
operators from the compatibility groups for $p \geq 3$.    Less essential is that the 
four generators are related by pairwise 
permutations of operators (hence the notation $\P4$).   The eigenstates have Bell
correlations between all pairs of particles, not just the chosen pairs as in $BB$ states.
So, unlike cluster or $BB$ states, the entanglement is shared equally among all four 
particles, but unlike GHZ states, the entanglement is robust.   One can perform 
measurements on any two particles, in any two different bases, and produce a
Bell state of the other two.
 
To show that the $\P4$ basis does not exhaust the possibilities for $p \geq 3$, we
mention another generator set involving cyclic permutations, ($ZXYW, XYWZ^{-1}, 
YWZX, WZ^{-1}XY$), where the $Z^{-1}$ factors are inserted for compatibility.   The 
corresponding basis could play a role similar to that of
$\P4$ in filling MUB complements for $p \geq 5$, although it turns out to be relatively 
inconsequential when $p = 3$.   In any case, since it would needlessly complicate 
the discussion without changing our conclusions, we exclude this example from the 
analysis.  

It is interesting to note in passing, that despite the differences in appearance among 
the generator sets of the four (five) totally entangled bases, the total numbers of 
$I_k$ factors appearing in their compatibility groups must be the same, namely 
4($p^2-1$), in accordance with Theorem II.   This has consequences for
stoichiometry, in particular for the standard distributions, and justifies classifying 
$BB$ bases as totally entangled.

Let us now turn to questions of stoichiometry.    While in previous cases we were 
able to deduce the allowed entanglement patterns from global constraints alone
(those involving total numbers of pure and entangled qupits in MUB complements), 
with $N \ge 4$ this is no longer the case.   The existence of multiple totally 
entangled basis types requires that we consider more microscopic constraints
associated with the distributions of $I_k$ factors, as was done for qubits in 
\Ref{Romero}.   To this end, we define a quantity that is capable of distinguishing 
among all MUB types under consideration.   

The ``$n$-body profile'' of a particular MUB is the distribution of $n$-body
operators ($n=1$, 2, ..., $N$) in its compatibility group, where (as implied earlier)
$n$-body operators are those with $N-n$ identity factors, $I_k$.  This distribution 
is normalized to the total number of operators in the compatibility group, $p^N-1$.    
Examples of $n$-body profiles are given in Table II, where we include the 
$N=2$ and 3 cases both for comparison with $N=4$, and also to show how 
global information is recovered.   The number of operators in each category, 
summed over all MUBs, must equal the numbers listed at the bottom of each 
column.  The latter represent the $n$-body profile of the set of all Pauli 
operators, and are thus independent of the particular MUB choices.  They 
are determined by generating all of the Pauli operators as expansions in the 
tensor products,
\be
   (I_1,Z_1,X_1,...,X_1Z_1^{(p-1)}) \otimes ...
   \otimes (I_N,Z_N,X_N,...,X_NZ_N^{(p-1)}).
\label{PauliIk}
\ee
Thus, the total number of $n$-body operators is $\big( {N \atop n} \big) (p-1)^n$, 
as shown.   The condition that the MUB sums equal these bottom lines, column 
by column, provides ($N-1$) independent constraint equations.  (In the 
exceptional case of $N=2$, both equations are independent.)
\begin{table}
$$
\vbox{
  \halign{
 \hfil # \hfil \quad & \hfil # \hfil \quad & \hfil # \hfil      \cr
     & \multispan{1 (a) $N=2$}          \cr
\noalign{\medskip}
\noalign{\hrule}
\noalign{\smallskip}
     & 1-body & 2-body                       \cr
\noalign{\smallskip}
\noalign{\hrule}
\noalign{\smallskip}
    $\Pi$ & 2($p-1$) & ($p-1)^2$      \cr
    $B$ & 0 & $p^2-1$                        \cr
\noalign{\smallskip}
\noalign{\hrule}
\noalign{\smallskip}
   all & 2($p^2-1$) & ($p^2-1)^2$     \cr
\noalign{\smallskip}
\noalign{\hrule}
\noalign{\bigskip}
}} 
\hskip1.5truecm
\vbox{
  \halign{
 \hfil # \hfil \quad & \hfil # \hfil \quad & \hfil # 
   \hfil \quad & \hfil #  \hfil                                        \cr
     &  \multispan{1 (b) $N=3$}             \cr
\noalign{\medskip}
\noalign{\hrule}
\noalign{\smallskip}
     & 1-body & 2-body & 3-body       \cr
\noalign{\smallskip}
\noalign{\hrule}
\noalign{\smallskip}
    $\Pi$ & 3($p-1$) & 3$(p-1)^2$ & ($p-1)^3$          \cr
    $SB$ & $p-1$ & $p^2-1$ & ($p-1)(p^2-1$)            \cr
    $G$ & 0 & 3($p-1$) & ($p-1)^2(p+2$)                     \cr
\noalign{\smallskip}
\noalign{\hrule}
\noalign{\smallskip}
  all & 3($p^2-1$) & 3($p^2-1)^2$ & ($p^2-1)^3$      \cr
\noalign{\smallskip}
\noalign{\hrule}
}}  
$$ 
\smallskip
%
$$
\vbox{
  \halign{
 \hfil # \hfil \quad & \hfil # \hfil \quad & \hfil # 
   \hfil \quad & \hfil # \hfil \quad & \hfil # \hfil     \cr
     & & \multispan{1 (c) $N=4$}    \cr
\noalign{\medskip}
\noalign{\hrule}
\noalign{\smallskip}
     & 1-body & 2-body & 3-body & 4-body     \cr
\noalign{\smallskip}
\noalign{\hrule}
\noalign{\smallskip}
    $\Pi$ & 4($p-1$) & 6($p-1)^2$ & 4($p-1)^3$ &  ($p-1)^4$           \cr
    $S^2B$ & 2($p-1$) & 2$p(p-1$) & 2($p-1)(p^2-1$) & ($p-1)^3(p+1$)      \cr
    $SG^{(3)}$ & $p-1$ & 3($p-1$) & ($p-1)^2(p+5$) & ($p-1)^3(p+2$)      \cr
    $BB$ & 0 & 2($p^2-1$) & 0 & ($p^2-1)^2$                                               \cr
    $G^{(4)}$ & 0 & 6($p-1$) & 4($p-1)(p-2$) & $p^4-4p^2+6p-3$         \cr
    $C^{(4)}$ & 0 & 2($p-1$) & 4$p(p-1$) & $p^4-4p^2+2p+1$           \cr
    $P^{(4)}$ & 0 & 0 & 4($p^2-1$) & $p^4-4p^2+3$                           \cr
\noalign{\smallskip}
\noalign{\hrule}
\noalign{\smallskip}
  all & 4($p^2-1$) & 6($p^2-1)^2$ & 4($p^2-1)^3$ & ($p^2-1)^4$     \cr
\noalign{\smallskip}
\noalign{\hrule}
\noalign{\smallskip}
\noalign{\hrule}
}}  
$$                                      
\caption{$n$-body profiles for all MUBs under discussion.   The bottom lines (``all'') 
are the $n$-body profiles of the set of all Pauli operators.}
\end{table}

It is immediately apparent from all of the first columns that the maximum number 
of $\Pi$ bases is always given by $p+1$, the number that defines the standard 
MUB complement.   Part (a) confirms that this is the only choice for $N=2$, and 
its second column then determines the number of Bell bases ($p^2-p$), in 
accordance with the required total number of MUBs.    Part (b) reproduces all 
$N=3$ results, as were summarized on Table I.    The three columns provide three 
equations, but only two are linearly independent:   The first column determines 
all possible combinations of $\Pi$ and $SB$ bases, and the second column then 
determines the number of $G^{(3)}$ bases, which is again consistent with the 
required total number of MUBs, $p^3+1$.   The third column provides no further 
constraint.

Proceeding to the case of $N=4$, the calculation of the $n$-body profiles for the
separable bases is straightforward, since their compatibility groups are tensor 
products of those whose profiles have already been calculated.    The new 
nonseparable bases require more thought.   We found that the more symmetrical 
generator sets listed in Eqs. \ref{genG}-\ref{genP}  were helpful in working out 
the profiles for general $p$.

One can see by inspection of Table II(c) that there is a qualitative difference between 
$N=4$ and the other cases.   Consider just the first 6 MUB types, which represent all 
possibilities for $p=2$.   Looking at the 3-body factors in column (iii) , we can 
see that as $p$ increases, the number of $\G4$ and/or $\C4$ MUBs would have 
to increase as $\sim p^4$ in order to satisfy just Eq. (iii).  But then they could not 
satisfy Eq. (ii), for they would produce too many two-body operators.   Clearly, 
one eventually needs a basis which, like $\P4$, has no two-body operators.   
This need makes itself felt already with $p=3$, and becomes urgent with $p=5$.   
With these differences in mind, let us consider the $p=2$, 3 and 5 cases 
sequentially, to show how the general picture evolves with increasing $p$. 

\centerline{{\bf four qubits}}
\begin{table}
$$
\vbox{
  \halign{
  \hfil # \hfil \quad & \hfil # \hfil \quad & \hfil # 
   \hfil \quad & \hfil # \hfil \quad & \hfil # \hfil     \cr
\noalign{\smallskip}
\noalign{\hrule}
\noalign{\smallskip}
     & 1-body & 2-body & 3-body & 4-body     \cr
\noalign{\smallskip}
\noalign{\hrule}
\noalign{\smallskip}
    $\Pi$ & 4 & 6 & 4 & 1                            \cr
    $S^2B$ & 2 & 4 & 6 & 3                       \cr
    $SG^{(3)}$ & 1 & 3 & 7 & 4                 \cr
    $BB$ & 0 & 6 & 0 & 9                           \cr 
    $G^{(4)}$ & 0 & 6 & 0 & 9                    \cr
    $C^{(4)}$ & 0 & 2 & 8 & 5                    \cr
\noalign{\medskip}
\noalign{\hrule}
\noalign{\medskip}
  all & 12 & 54 & 108 & 81                       \cr
\noalign{\smallskip}
\noalign{\hrule}
}}
$$                                      
\caption{Specific $n$-body profile for four qubits.}
\end{table}

The $n$-body profiles for $p=2$ are shown on Table III.  To explore 
stoichiometries, consider the three equations 
(i, ii, and iii) represented by the first three columns, respectively.   Equation (i), 
by itself,  determines all possible combinations of the first three MUB types, 
\be 
  4N(\Pi)+2N(S^2B)+N(SG^{(3)}) = 12.
\label{N4p21}
\ee
Next, notice that we can isolate the $BB$ and $G^{(4)}$ MUBs because of  their 
simple profiles.   Indeed, by simply adding (i) and (iii) we obtain the sum of all
{\it other} MUBs,
\be
    N(\Pi) + N(S^2B) + N(SG^{(3)}) + N(C^{(4)}) = 15.
 \label{N4p22}
 \ee
Since there are 17 MUBs in total we know immediately that
\be
   N(BB) + N(G^{(4)}) = 2,
\label{N4p23}
\ee
a result which also follows from 2(ii) + (iii) - (i), which reproduces the total number.   

There are 16 ways to satisfy \Eq{N4p21}, with $N(\C4$) determined in each case by
\Eq{N4p22}.   For each of these combinations, there are 3 ways to satisfy \Eq{N4p23},
for a total of 48 possible MUB distributions.     To illustrate the range, a standard and 
a nonstandard distribution are shown in the two leftmost columns of Table IV.  These
examples are chosen to show the maximum and minimum numbers of the new
(nonseparable) $\C4$ MUBs, which make up a majority of MUBs in 30 of the 48 
possible distributions.   The dominance of $\C4$ MUBs is related to the large number
of 3-body operators in their profile.  Similar complements were found in \Ref{Romero} 
through an explicit construction, except that the $G^{(4)}$ MUBs were not produced, 
so that 16 combinations were obtained with 2 $BB$ MUBs present in all of them.   

\begin{table}
$$
\vbox{
  \halign{  
  \hfil # \hfil \quad & \hfil # \hfil \quad & \hfil # 
   \hfil \quad & \hfil # \hfil  \quad & \hfil # \hfil \quad & \hfil # 
   \hfil \quad & \hfil # \hfil                                                                      \cr
\noalign{\smallskip}
\noalign{\hrule}
\noalign{\smallskip}
     & \multispan{2 ($p=2$) \hfil} & \multispan{2 ($p=3$) \hfil} & 
     \multispan{2 ($p=5$) \hfil}     \cr
\noalign{\smallskip}
\noalign{\hrule}
\noalign{\smallskip}
    $\Pi$ & 3 & 0 & 4 & 0 & 6 & 0                                   \cr
    $SG^{(3)}$ & 0 & 12 & 0 & 16 & 0 & 24                 \cr
    $BB$ & 2 & 2 & 0 & 0 & 0 & 0                                    \cr 
    $C^{(4)}$ & 12 & 3 & 72 & 66 & 360 & 396          \cr
    $\P4$ & -- & -- & 6 & 0 & 260 & 206                       \cr
\noalign{\medskip}
\noalign{\hrule}
\noalign{\medskip}
  all & 17 & 17 & 82 & 82 & 626 & 626                        \cr
\noalign{\smallskip}
\noalign{\hrule}
}}
$$                                      
\caption{Examples of MUB distributions for four qupits with $p=2$, 3, and 5.
First, third, and fifth columns show standard distributions that maximize the
numbers of totally entangled bases, while even columns show nonstandard
distributions that minimize this number.    Examples are chosen to minimize 
the number of $\P4$ MUBs in all of the $p=3$ and 5 cases.}
\end{table}

\centerline{{\bf four qutrits}}

The $n$-body profiles for the $p=3$ case are shown in Table V.   Again,
the first column restricts the combinations of the first three MUB types,
\be
   4N(\Pi) + 2N(S^2B) + N(SG^{(3)}) = 16.
\label{N4p31}
\ee
The first and second columns together [(ii)$-$3(i)] restrict other combinations,
\be 
   4N(BB) + 3N(G^{(4)}) + N(C^{(4)}) = 72,
\label{N4p32}
\ee
and the inclusion of the third column [(iii)$+$2(ii)$-$6(i)] yields the 
total MUB count,
\be
   N(\Pi) + ... + N(P^{(4)}) = 82.
\label{N4p33}
\ee
\begin{table}
$$
\vbox{
  \halign{  
  \hfil # \hfil  \quad & \hfil # \hfil \quad & \hfil # 
   \hfil \quad & \hfil # \hfil                                    \cr
     & \multispan{1 (a) $p=3$}                             \cr
\noalign{\smallskip}
\noalign{\hrule}
\noalign{\smallskip}
     & 1-body & 2-body & 3-body        \cr
\noalign{\smallskip}
\noalign{\hrule}
\noalign{\smallskip}
    $\Pi$ & 8 & 24 & 32                                \cr
    $S^2B$ & 4 & 12 & 32                            \cr
    $SG^{(3)}$ & 2 & 6 & 32                         \cr
    $BB$ & 0 & 16 & 0                                   \cr
    $G^{(4)}$ & 0 & 12 & 8                            \cr
    $C^{(4)}$ & 0 & 4 & 24                         \cr
    $P^{(4)}$ & 0 & 0 & 32                      \cr
\noalign{\medskip}
\noalign{\hrule}
\noalign{\medskip}
  all & 32 & 384 & 2048                        \cr
\noalign{\smallskip}
\noalign{\hrule}
}}
\hskip1truecm
\vbox{
  \halign{  
  \hfil # \hfil  \quad & \hfil # \hfil \quad & \hfil # 
   \hfil \quad & \hfil # \hfil                                      \cr
     & \multispan{1 (b) $p=5$}                             \cr
\noalign{\smallskip}
\noalign{\hrule}
\noalign{\smallskip}
      & 1-body & 2-body & 3-body                        \cr
\noalign{\smallskip}
\noalign{\hrule}
\noalign{\smallskip}
     $\Pi$ & 16 & 96 & 256                                       \cr
     $S^2B$ & 8 & 40 & 192                                     \cr
     $SG^{(3)}$ & 4 & 12 & 160                               \cr
     $BB$ & 0 & 48 & 0                                             \cr
     $G^{(4)}$ & 0 & 24 & 48                                    \cr
     $C^{(4)}$ & 0 & 8 & 80                                      \cr
     $P^{(4)}$ & 0 & 0 & 96                                      \cr
\noalign{\medskip}
\noalign{\hrule}
\noalign{\medskip}
    all & 96 & 3456 & 55296                                    \cr
\noalign{\smallskip}
\noalign{\hrule}
}}
$$                                      
\caption{The $n$-body profiles for four qutrits and four ququints.}
\end{table}
There are 25 combinations of the first three MUB types that satisfy \Eq{N4p31}, 
as compared with 16 such combinations in the qubit case.   But, in the absence
of $\P4$ MUBs, one cannot solve both \Eqs{N4p32}{N4p33} for all of these
combinations, and we  find a total of only 11 MUB distributions.   To trace the
reasons, we subtract \Eq{N4p32} from \ref{N4p33} and solve for $\P4$:
\be
   N(P^{(4)}) = 10 + 3N(BB) + 2N(G^{(4)}) - [N(\Pi) + N(SSB) + N(SG^{(3)})].
\label{N4p34}
\ee
Without $\P4$ MUBs the left side vanishes, and there can be solutions only if
the quantity in square brackets is 10 or larger.   This condition fails for the 
standard distribution, for which this number is $N(\Pi)=4$.  In this case, the 
minimum number of $\P4$ MUBs is 6, as shown on Table IV.  The other entry 
maximizes the quantity in square brackets at $N(SG^{(3)})=16$.  In both entries 
we then minimize the number of $\P4$ MUBs by maximizing the number of $\C4$ 
MUBs.   One can increase the number of $\P4$ MUBs over these minima by 
adding $BB$ and/or $\G4$ and subtracting $\C4$ MUBs.   Although its numbers 
can be small, the $\P4$ MUBs play a critical role in maintaining the balance of 
3-body operators (Table V) at no cost in two-body operators. 

With $\P4$ MUBs included, the multiplicity of each of the 25 solutions of 
\Eq{N4p31} is large (we estimate more than 200), for a total of probably more 
than 5000 solutions.  We cannot argue that all of these solutions represent 
realizable MUB distributions, because we cannot rule out the possibility of 
more subtle constraints.   Such concerns are beyond the scope of the present
paper.

\centerline{{\bf four ququints}}

Again consulting Table V for the $p=5$ case, it is striking to see how three 
simple equations can again emerge from appropriate combinations.   The first 
column gives us directly
\be
   4N(\Pi) + 2N(S^2B) + N(SG^{(3)}) = 24,
\label{N4p51}
\ee
the combination  [(iii)$+$8(ii)$-$64(i)] relates the other four quantities,
\be 
   8N(BB) + 5N(G^{(4)}) + 3N(C^{(4)}) + 2N(\P4) = 1600,
\label{N4p52}
\ee
and still another combination  [(iii)$+$2(ii)$-$22(i)] yields the total MUB count,
\be
   N(\Pi) + ... + N(P^{(4)}) = 626.
\label{N4p53}
\ee
There are 49 combinations of the first three MUB types that satisfy \Eq{N4p51},
but in the absence of the $\P4$ MUBs, {\it none} of these admits solutions of 
\Eqs{N4p52}{N4p53}.  To see how this situation arises, solve the latter
two equations for $N(\P4)$ while eliminating $N(\C4)$:
\be
   N(P^{(4)}) = 278 + 5N(BB) + 2N(G^{(4)}) - 3[N(\Pi)+N(SG^{(3)})+N(SSB)].
\label{N4p54}
\ee
The quantity in square brackets has minimum and maximum values of 6 (the
standard distribution) and 24, as shown on Table V,  corresponding to lower 
bounds on $N(\P4)$ of 260 and 206, respectively.  The latter is the absolute
minimum number of $\P4$ MUBs in any full complement.   Again, one can add
$\P4$ MUBs by removing $\C4$ and adding $BB$ and/or $\G4$ MUBs, so that
$\P4$ can be the majority MUB type in some complements.   While $\P4$ is 
critical for both $p=3$ and 5 cases, it  plays a considerably more dominant role 
here.   The underlying reason is that the ratio of the numbers of 3-body to 2-body 
operators increases considerably in going from $p=3$ to 5, as shown in Table V.

We estimate the total number of solutions of Eqs. \ref{N4p51}-\ref{N4p53} to
be in excess of $10^6$, but again, we cannot argue that all such solutions 
represent realizable MUB distributions, or provide a revised estimate, without 
a further study of possible constraints.   

The examples of this section have shown us that with every step in $N$, and 
with some steps in $p$, full complements require not only those MUB types 
generated from smaller systems, but also new, nonseparable MUB types that 
exhibit new entanglement characteristics inaccessible to smaller systems.   
In the step from $N=2$ to 3, $\GG3$ MUBs are required for the standard 
distribution, although a nonstandard distribution ($SB$ only) is possible with 
$p=2$.   With the step to $p=3$, no MUB complement exists without $\GG3$.  
In the step to $N=4$, the $\C4$ MUBs are indespensible to all MUB distributions 
with $p=2$.  With the step to $p=3$, the new $\P4$ MUBs become possible, and 
they in turn make possible the standard distribution.   At $p=5$, the $\P4$ MUBs 
become indispensable to all distributions.

Projecting to larger systems, the distinguishing feature of the $\P4$ generator 
set is that a different (noncommuting) Pauli matrix factor is associated with each 
qupit.  The number of such factors in general is $p+1$, and when this is equal 
to the number of qupits, a new type of entanglement becomes possible.   Thus
we predict that when $N$ is equal to any prime plus 1, then that prime 
($p_N^{} = N-1$) is a critical value for the emergence of new entangled states
as $p$ is increased at fixed N.   These states should play critical roles 
in filling MUB complements for $p$ equal to or slightly greater than $p_N$.

\section{Conclusions and Open Questions}

We have exploited the connections between MUBs and Pauli operators 
to develop a general framework for investigating both the entanglement
properties of individual MUBs, and the combinations of such MUBs that can
be found in full complements.  We began by proving general theorems
regarding MUBs as eigenbases of Pauli operators:  We showed that the 
purities of individual qupits in such eigenbases must be either 0 or 1, that the
purity alone dictates the distribution of Pauli matrix factors (including $I_k$) 
in the compatibility groups of these MUBs, 
and that every qupit must adopt these special purities the same number 
of times within any MUB complement:  ($p+1$) times pure, and ($p^N-p$) 
times totally entangled.   An immediate corollary is that one may have at most 
$p+1$ product MUBs in a full complement, and when one does, all remaining 
MUBs must be totally entangled.  This defines the standard distribution. 

Armed  with these theorems and the general properties of Pauli operators, 
one quickly obtains more specific results:    When $N=2$, only product and 
generalized Bell bases are possible, for any $p$, and the standard MUB 
distribution is inevitable.   With $N=3$, the unique totally entangled bases 
are generalized GHZ bases, but a third MUB type becomes possible, namely
separable-Bell bases.  This makes possible $p+2$ distinct MUB distributions.  
With $N=4$ and $p=2$ there are six MUB types, including two nonseparable 
bases and a third ($BB$) that is separable but totally entangled.  There are 48 
possible MUB distributions, with cluster bases making up the majority of MUBs 
in most of these.  With $N=4$ and larger $p$, further MUB types exist, and at 
least one such MUB type ($\P4$) is essential to forming a standard MUB 
complement with $p=3$, and to forming {\it any} MUB complement with $p=5$.   

Several results have emerged in the course of working the above examples, and 
it seems useful to synthesize these in one place:  (1)  A MUB can exist in any
separation pattern.  (2)  All states in a particular MUB have common separation 
and entanglement patterns - the generator set contains all information about 
the nature of the entanglement, while the eigenvalues specify the states. 
(3) Compatibility groups of separable bases are tensor products of those of the 
nonseparable constituent bases.  It follows that (4) within nonseparable 
groupings of qupits, those with two qupits must be in generalized Bell states, 
those with three qupits - generalized GHZ states.  Those with 4 qupits have the 
same broader array of options available to 4-qupit systems. 

Perhaps the most important lesson to be drawn from the present examples is 
that, although it is easy to construct MUBs from those found at lower $N$, either 
as tensor products, or as larger-$N$ counterparts such as $G^{(N)}$, the more 
interesting challenge is to find the new nonseparable MUB types, with no 
counterparts at smaller $N$ (or sometimes $p$), that make full complements 
possible.  It may be a general feature that such MUBs tend to dominate MUB
distributions near the $N$ and $p$ values where they first emerge, 
only to be superceeded by other MUB types as the system size increases.   In 
this sense, every $N$ is critical, but not every $p$.   We predict that when $N$ 
is a prime plus 1 (\eg, $N=6$, 8,...), there will be a critical value, $p_N = N-1$, 
for the introduction of new entangled states that will play critical roles in MUB
distributions.

Closing thoughts on the existence question:  The intimate connection between 
MUBs and entanglement for $N \geq 2$ highlights the way in which all known 
MUB complements take advantage of the symmetry associated with equivalent 
parts, unique to dimensions $p^N$ \cite{Durt3}.  Theorem I, which makes no 
reference to dimension, cannot hold in (at least some) composite dimensions:  In 
the simplest counterexample, 6 dimensions, the qubit can be totally entangled, 
but the qutrit cannot be.   Yet, the import of Theorem I is that for all known MUB 
complements, there exists a factorization into parts (represented by some set of 
generalized Pauli operators), in terms of which Theorem I (and the others) hold.  
Thus, in addition to the existence question in composite dimensions, there is 
also an existence question in $p^N$ dimensions - Do MUB complements exist
that violate Theorem I?   Perhaps entanglement considerations such as the 
present ones will help in answering these persistent questions.

\bigskip
\centerline{{\bf ACKNOWLEDGEMENTS}}

I would like to thank Bill Wootters and Winton Brown for many stimulating 
discussions, and I would like to thank the James Franck Institute for its hospitality 
during the time when this work was completed.

\bigskip
\centerline{{\bf APPENDIX}}

First, we write general definitions of $\G4$ and $\C4$ bases in terms of generator 
sets, and then we show that these are the only possible nonseparable bases for 
four qubits.

To generalization \Eqs{genG}{genC} with maximum transparency, we follow the
alternative forms written in the text and define $\G4$ bases by the generator set 
\be
    \G4 = (ABII,~AICI,~AIID,~STUV),
\label{AgenG}
\ee
where every four-body factor must differ from its two-body counterpart ($S \neq A$,
etc.).   The two-body operators provide the special product basis for the $p$-term 
expansion.   The generalization to $N$-qupit GHZ states is apparent.

The $\C4$ bases are best defined in a similar way, although a bit less transparently
because there are only two independent two-body operators,
\be
   \C4 = (AICI,~IBID,~SBUI,~ITCV).
\label{AgenC}
\ee
Again, the two-body operators provide a product basis for the expansion, which in 
this case requires $p^2$ terms.   Individual factors in the three-body operators must 
differ from corresponding factors in the two-body operators, except where their 
equality is explicit.   

It should be noted that there are two variations on the $\C4$ generator set, 
corresponding to the other ways of pairing the two-body factors.   One such variation 
is ($ABII,~IICD,~STCI,~IBUV$).   The three alternatives are mathematically 
equivalent, although one can make a physical distinction based on entanglement 
links between pairs.   The stronger entanglement links in a system represented by  
\Eq{AgenC} are between neighbors in the sequence (1-2-3-4-1).   In the variation 
given above, the sequence is (1-3-2-4-1), and in the other possible variation it is 
(1-2-4-3-1).  The various possibilities are not unphysical, as one can imagine unlike
particles with tetrahedral coordination. 

\centerline{{\bf Completeness for qubits}}

We now argue that the two bases defined above are the only nonseparable 
options for qubits.   We first argue that both are the unique nonseparable 
representatives of their respective $n$-body profiles as shown on Table III.   
We then show that other profiles cannot exist for qubits.   

It is straightforward to verify that the $\G4$ generators produce six two-body 
operators involving the same factors, $A$ - $D$.   These exhaust all 12 of the 
available $I_k$ factors (Theorem II demands three $I_k$ factors per qubit), 
so that all remaining operators are 4-body operators, as shown in the table.
The only other basis that can share this profile is $BB$, in which the two-body
operators have factors that do not commute individually.   But, by virtue of this fact,
the $BB$ basis has four independent two-body operators, so that these can 
compose the generator set.   The separability of the basis is then obvious. 

Turning to the $\C4$ case, it is straightforward to show that the generators of
\Eq{AgenC} produce no further two-body operators beyond the two shown, so that 
remaining $I_k$ factors must appear with the 8 three-body operators.   One might 
wonder whether a different basis could be found with the same profile by using 
four-body generators in place of the three-body generators.   The answer is no - 
It is easy show that this would generate only $\G4$ or $BB$ bases, depending 
upon whether one of the four-body generators shares factors with one of the 
two-body generators.

The remaining point is to rule out other four-qubit profiles.  It suffices to consider
just the two-body operators, whose maximum number is six.   We will show that the 
numbers 4 and 0 are impossible for qubits.  The former case is very simple - it is 
impossible to find four commuting two-body operators that do not generate two more 
(and these will immediately identify themselves as belonging to either a $\G4$ or 
$BB$ compatibility group).    As to the latter case, assume that there are no two-body
operators.  Then all 12 of the $I_k$ factors must appear in one-body operators, 
making three appearances on each qubit (Theorem II).   Consider any two 
of these operators that have their $I_k$ factors on the same qubit.   Commutativity 
demands that they have exactly one other factor in common, so that their product 
is a two-body operator.  This forms a contradiction and shows that there is no
profile without two-body operators.

To briefly summarize the results of this appendix, the 6 MUB types listed on Table III 
exhaust the possibilities for four qubits.    The 5 corresponding $N$-body profiles are
also exhaustive; in particular, a $\P4$-like profile does not exist for qubits.      

\begin{thebibliography}{99}
\bibitem{transform} One may think of this as an $N$-dimensional discrete Fourier 
   transform with $p$ points in each dimension, or as a Fourier transform over a 
   Galois field, with $n$ and $k$ as Galois field variables.
\bibitem{Pauligroup} The one-qupit Pauli groups may be written 
   $U_i \otimes (1,\om,...,\om^{p-1})$ for odd $p$, and $U_i \otimes (\pm 1, \pm i)$ for
   $p=2$, where $U_i$ represents all of the $p^2$ Pauli matrix factors.  These are
   special cases of the discussion following \Eq{Pauli2}.
\bibitem{Wiesniak2} This result was derived in another way in \Ref{Wiesniak}.
\bibitem{pure}  A one-body Pauli operator transforms only a single qupit
   (for example, $IU$ in a 2-qupit system and $VII$ in a 3-qupit system).  Such an
   operator can have no totally entangled eigenstates, because in any eigenstate the 
   transformed qupit must be pure.
\bibitem{Durt3} This symmetry is referred to in the concluding paragraph (p. 91)
    of \Ref{Durt}.
\end{thebibliography}
\end{document}